\documentclass[12pt]{spieman}  

\usepackage{amsmath,amsfonts,amssymb}
\usepackage{graphicx}
\usepackage{setspace}
\usepackage[subfigure]{tocloft}

\usepackage{soul}
\usepackage[export]{adjustbox}
\usepackage{subfigure}
\usepackage{hyperref}

\usepackage{caption}
\usepackage[leftcaption]{sidecap}
\usepackage{soul}
\usepackage{newfloat}
\usepackage{xr}
\usepackage{indentfirst}
\usepackage{hyperref}
\usepackage{tikz,xcolor,hyperref}
\usepackage{svg}
\usepackage{svg-extract}
\svgsetup{clean}

\usepackage{blindtext}
\usepackage{titlesec}
\usepackage{setspace}
\usepackage{siunitx}
\DeclareSIUnit{\molar}{M}
\DeclareSIUnit{\wpv}{w/v}

\DeclareSIUnit{\calorie}{cal}
\DeclareSIUnit{\Calorie}{\kilo\calorie}

\usepackage{etoolbox}
\makeatletter
\patchcmd{\@makefntext}{\hss}{}{}{} 
\makeatother

\definecolor{myBlue}{rgb}{0.1,0.1,0.7}

\title{{Unleashing Urease Protein Dynamics with Metamaterial Optical Tweezers}}

\author[a,b,*,\textdagger]{Domna G. Kotsifaki}
\author[a,\textdagger]{Viet Giang Truong}
\author[c,d]{Mirco Dindo}
\author[e]{Frank Cichos}
\author[c]{Paola Laurino}
\author[a,*]{{S\'{i}le Nic Chormaic}}

\affil[a]{Light-Matter Interactions for Quantum Technologies Unit, Okinawa Institute of Science and Technology Graduate University, Onna, Okinawa 904-0495, Japan}
\affil[b]{Photonics Lab, Division of Natural and Applied Sciences, Duke Kunshan University, 8 Duke Ave, Kunshan, Jiangsu Province 215316, China}
\affil[c]{Protein Engineering and Evolution Unit, Okinawa Institute of Science and Technology Graduate University, Onna, Okinawa 904-0495, Japan}
\affil[d]{Department of Medicine and Surgery, Università degli Studi di Perugia, Piazza Università,Perugia 06123, Italy}
\affil[e]{Peter Debye Institute for Soft Matter Physics, Molecular Nanophotonics Group, Universität Leipzig, 04103, Leipzig, Germany}

\cftpagenumbersoff{figure}
\cftpagenumbersoff{table}
\begin{document}
\maketitle

\begin{abstract}
Investigating the dynamics of single biomolecules is essential for unlocking new frontiers in biophysics and medicine. Here, we present a transformative approach using metamaterial optical tweezers to trap and study individual urease molecules - an enzyme that catalyzes urea hydrolysis and serves as a key biomarker for pathogenic infections. By
generating thermally induced drifts and rapid hydrodynamic flows with optically induced local temperature fields at the metamaterial/water interface, we achieve unprecedented guidance and control over single urease molecules at low trapping laser intensities of 0.13~mW/\si{\micro\meter\squared} facilitating their precise delivery to plasmonic hotspots. The interplay of thermophoresis, thermo-osmosis and depletion forces shapes the trapping potential, enabling real-time observation of ureases` conformational changes and interactions with the environment. This innovative approach not only enhances our understanding of enzyme dynamics but also paves the way for label-free characterization of biomolecules, ushering in a new era in life sciences and nanotechnology.

\end{abstract}


{\noindent \footnotesize\textbf{*}Domna G. Kotsifaki,  \linkable{domna.kotsifaki@dukekunshan.edu.cn} }
{\noindent \footnotesize\textbf{*}S\'{i}le {Nic Chormaic}, \linkable{sile.nicchormaic@oist.jp} }

{\noindent \footnotesize\textbf{\textdagger}{These authors contributed equally to this work.}

\begin{spacing}{2}   

\section{Introduction}
In biomedical research and disease diagnosis, identifying single biomolecules in solution is essential.
Several advanced techniques and methods are used to meet this goal~\cite{Farka}. For example, fluorescence resonance energy transfer (FRET) can be used to detect conformational changes by measuring the transfer of energy between two fluorophores attached to interacting biomolecules~\cite{Silas}. However, repeated excitation of fluorophores can lead to photobleaching, reducing the intensity of fluorescence signals over time, which may affect the accuracy of such measurements~\cite{Silas}.
Since optical tweezers~\cite{Ashkin} can measure forces and torques with high accuracy and temporal resolution, they could be considered as a preferred method for single-particle characterization in many biophysical applications~\cite{Bustamante}.
Despite the wide range of applications~\cite{Volpe_2023}, optical tweezers face certain challenges, particularly when it comes to tethering and labeling molecules.
To overcome the limitations of conventional optical tweezers for the detection of single molecules, plasmonic optical tweezers (POT) were developed~\cite{Novotny,Pang, Kotsifaki1,Bouloumis_2021}.
Among various plasmonic nanostructure designs, metamaterials have emerged as particularly promising due to their unique electromagnetic properties that surpass those of natural materials~\cite{Boris}. These engineered structures provide large active areas crucial for studying interactions between resonating metamolecules and biological entities, enabling detection of viruses~\cite{Ahmadivand}, bacteria~\cite{KotsifakiBOE}, and other biomolecules~\cite{Wu}. Of particular interest are metamaterials supporting Fano-like resonances, which exhibit narrow spectral windows where scattering is suppressed and absorption is enhanced through interference between super-radiant and subradiant plasmonic modes~\cite{Papasimakis}. The high geometric sensitivity of these resonances, combined with light confinement to sub-diffraction volumes, makes such structures ideal for applications ranging from ultrasensitive biomolecule detection~\cite{Wu} to efficient nanoparticle trapping at low trapping laser intensities, $I_{trap}$,~\cite{Kotsifaki1, DomnaAPL,Bouloumis}.

Meanwhile, the energy losses in plasmonic substrates and the heat generated at the nanoscale offer significant advantages for a broad range of applications, including photothermally assisted plasmonic sensing~\cite{Baffou_Nat}. This localized heating can induce changes in the refractive index of the surrounding medium which are detectable and can be correlated with analyte concentrations~\cite{Zijlstra_NatNanotech}.
Beyond sensing applications, the nanoscale heating generated in the liquids can cause thermophoresis, driving the suspended particles due to temperature gradients~\cite{Wurger_2010}. Besides thermophoresis, the temperature gradient can also trigger companion effects like temperature-generated hydrodynamic flows~\cite{Bregulla}, thermo-viscous flows~\cite{Weinert} or concentration gradients~\cite{Sano,Jiang_2024} that can facilitate the manipulation of suspended particles well beyond the optical trapping region. The combination of these effects serves as a powerful approach for nanoscale manipulation and opens new avenues for studying single-molecule interactions or understanding the dynamics of macromolecules in temperature fields~\cite{Franz_2019}.

Here, we expanded the capabilities of metamaterial optical tweezers by introducing a thermo-plasmonic-assisted concept that enables the dynamic optical manipulation of sub-\SI{10}{\nano\meter} objects at positions removed from the high-intensity laser focus. Experimental demonstration of trapping was performed using urease biomolecules that have a hydrodynamic radius of \SI{7.4}{\nano\meter}, and \SI{20}{\nano\meter} polystyrene (PS) particles as a control solution, suspended in polyethylene glycol(PEG)/heavy water. Urease is a nickel-containing enzyme that catalyzes the hydrolysis of urea to form ammonia and carbon dioxide in some bacteria, fungi, algae, and plants~\cite{Follmer}. Healthy human organisms do not contain urease; however,~\textit{Helicobacter pylori}, a bacterial pathogen of the human stomach and a class I carcinogen~\cite{Damien}, secretes urease. In addition,~\textit{Canavalia juncea} urease (Jack bean) is widely used in biosensors for urea determination in biological fluids~\cite{Pundir}. PEG is a highly soluble polymer in water and has been used to modify biomolecules and to create more stable drug delivery systems for anticancer therapies~\cite{Gupta}. We observed that a single urease molecule can be delivered to, and trapped in, the nano-aperture of an asymmetric, split-ring (ASR) metamaterial using $I_{trap}$ as low as \SI{0.13}{\milli\watt\per\square\micro\meter}. By increasing the $I_{trap}$, the urease can be trapped at positions removed from the laser focus,~\textit{i.e.}, the center of the metamaterial. This minimizes phototoxicity and thermal stress induced by the laser while simultaneously enabling a real-time study of dynamic conformation changes.
By calculating the probability density function (PDF) of the trapped urease for several $I_{trap}$, we obtain valuable information about its dynamics such as position, orientation, and conformation states. Given these insights, our approach—enabling non-invasive trapping and manipulation of individual biomolecules at the single-particle level—holds great promise for advancing research in the life sciences.

\section{Results}
\noindent \textbf{Thermodynamic manipulation of urease in PEG$_{6000}$/Water: Schematic illustration, heat transfer simulation, and thermodynamic force analysis}\

A schematic of the experimental setup used to deliver, trap, and manipulate a single urease enzyme is shown in Figure~\ref{Fig.1}(a). The metamaterial consists of an array of 15 $\times$ 16 asymmetric split-ring (ASR) metamolecules with a periodicity of 400~\si{\nano\meter} milled on a 50~\si{\nano\meter} gold film (Figure~\ref{Fig.1}(b)). A solution of urease in PEG$_{6000}$/D$_{2}$O water was prepared. Additionally, a solution of polystyrene (PS) particles with a mean diameter of 20~\si{\nano\meter} in PEG$_{6000}$/D$_{2}$O water was used as a control.

\begin{figure}[h!]
\centering
\includegraphics[width=1\textwidth]{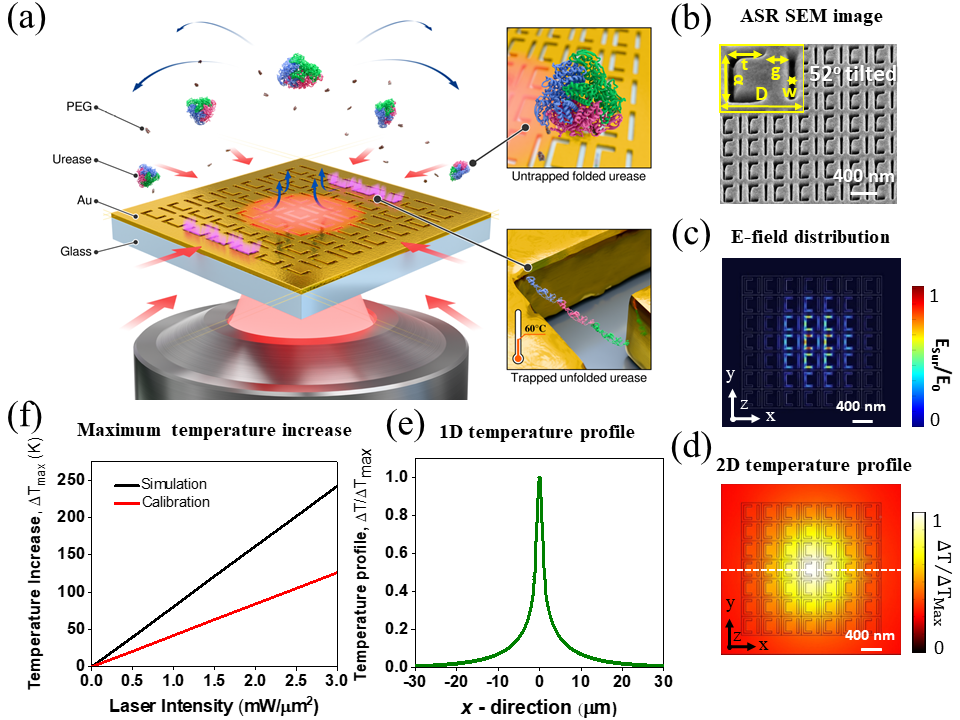}
\setlength\abovecaptionskip{0pt}
\caption{\label{Fig.1} \textbf{Description of the thermodynamic trapping concept and heat transfer simulations of the ASR metamaterial}. (a) A graphical visualization of the induced thermo-osmotic flows (see Figure~S6 and SI-1.3) to deliver and trap the enzyme in the excited plasmonic thermal hotspot is shown. The urease is made up of a single-chain polypeptide containing 840 amino acid residues~\cite{BALASUBRAMANIAN2010274}. Because urease's conformation may change when trapped, the enzyme has been drawn as a modified structure within the ASR metamolecule compared with its structure in a free solution (see insets). (b) Scanning electron microscopy (SEM) image of a 7~$\times$~7 array of the asymmetric split-ring (ASR) metamolecules fabricated from a 50~\si{\nano\meter} thin gold film, taken at an angle of 52\si{\degree} to the surface normal. The inset shows a metamolecule unit with feature dimensions: D = 400$~\pm~2$ \si{\nano\meter}, vertical slit $\alpha$ = 310~$~\pm~3$ \si{\nano\meter}, horizontal slit t = 165~$~\pm~3$ \si{\nano\meter}, gap g = 100~$~\pm~3$ \si{\nano\meter}, and slit width w = 44~$~\pm~2$ \si{\nano\meter}. (c) Spatial~\textit{E}-field distribution for an on-resonant excitation wavelength, $\lambda_{t}$ = 937~\si{\nano\meter}, polarized parallel to the \textit{x}-direction at the metamaterial surface. (d) Relative temperature profile generated in the \textit{xy}-plane at the metamaterial/water interface. (e) \textit{x}-direction cut in Figure~\ref{Fig.1}(d) (dashed white line). (f) Simulated temperature increase (black line), $\Delta T_{max}$, as a function of $I_{trap}$ and its calibration from the experimental absorption cross-section (see Figure~S3 and SI-1.2). $\Delta T_{max}$ was extracted at the center focal point of the incident Gaussian beam, \textit{i.e.}, 0~\si{\micro\meter} in the \textit{x}-direction.
}
\end{figure}

The~\textit{\textbf{E}}-field distribution was calculated using the entire 3D electromagnetic (EM) domain, as shown in Figure~\ref{Fig.1}(c) (see Figure~S1(a) and SI-S1.1). Next, once the~\textit{\textbf{E}}-field distribution is calculated, the heat source density, $q_{i}$(\textbf{r}), and the temperature distribution, T$(\textbf{r})$, can be then numerically solved using the heat transfer (HT) module (SI-S1.2). Figure~\ref{Fig.1}(d) shows a 2D profile of the relative temperature distribution, defined as \textit{ $\Delta$T = T$(\textbf{r})$ – $T_{0}$}, where \textit{T$_{0}$} = 293.15 $\si{\kelvin}$ is the ambient temperature, generated at the ASR metamolecules surface using the coupled finite element EM and HT modules (see Figure~S2 and SI-S1.2). Figure~\ref{Fig.1}(e) shows the corresponding 1D relative temperature distribution line plotted along the $\textit{x}$-direction through the middle of the array (dashed white line in Figure~\ref{Fig.1}(d)). We observe that at the center of the metamaterial, the strong heat source density leads to a temperature increase, while far from the center at 30~\si{\micro\meter}, the heat source is weaker and the temperature approaches the baseline ambient value, \textit{T$_{0}$}.

Under resonant illumination, $\lambda_t$ = 937~\si{\nano\meter}, a large portion (or most) of the incident photon energy is converted into collective electron oscillations within the ASR metamolecule cavities, leading to resistive heating linearly proportional to the absorption cross-section (see SI-S1.2). As a result, the collective heating from multiple ASR metamolecules raises the absolute temperature, T(\textbf{r}), and the heat spreads across the gold/water interface due to the high thermal conductivity of the thin gold film. To estimate the theoretical values of the absolute temperatures, we first simulated the wavelength-dependent absorption cross-section of the ASR metamaterial (Figures~S1(c) and (d) as well as SI-1.1). Then, we calculated the maximum temperature increase, $\Delta$$T_{max}$, in the center of the metamaterial, as shown in Figure~\ref{Fig.1}(e) (black line). Next, the amplitude of the theoretical absorption cross-sections is monitored by the experimental values, which are identified from the measured absorption spectrum (Figures~S1(c) (bottom) and S1(d)). Using these calibration values~\cite{Kotsifaki1,Bouloumis}, we recalculated $\Delta$$T_{max}$ (red line in Figure~\ref{Fig.1}(f) and Figure~S3 (SI-S1.2)), and observe that the temperature increment at the center of the metamaterial reaches $\Delta T$ = 125$\,\si{\kelvin}$ for $I_{trap}$ of 3~\si{\milli\watt\per\micro\meter\squared}.

\begin{figure}[]
\centering
\includegraphics[width=1.0\textwidth]{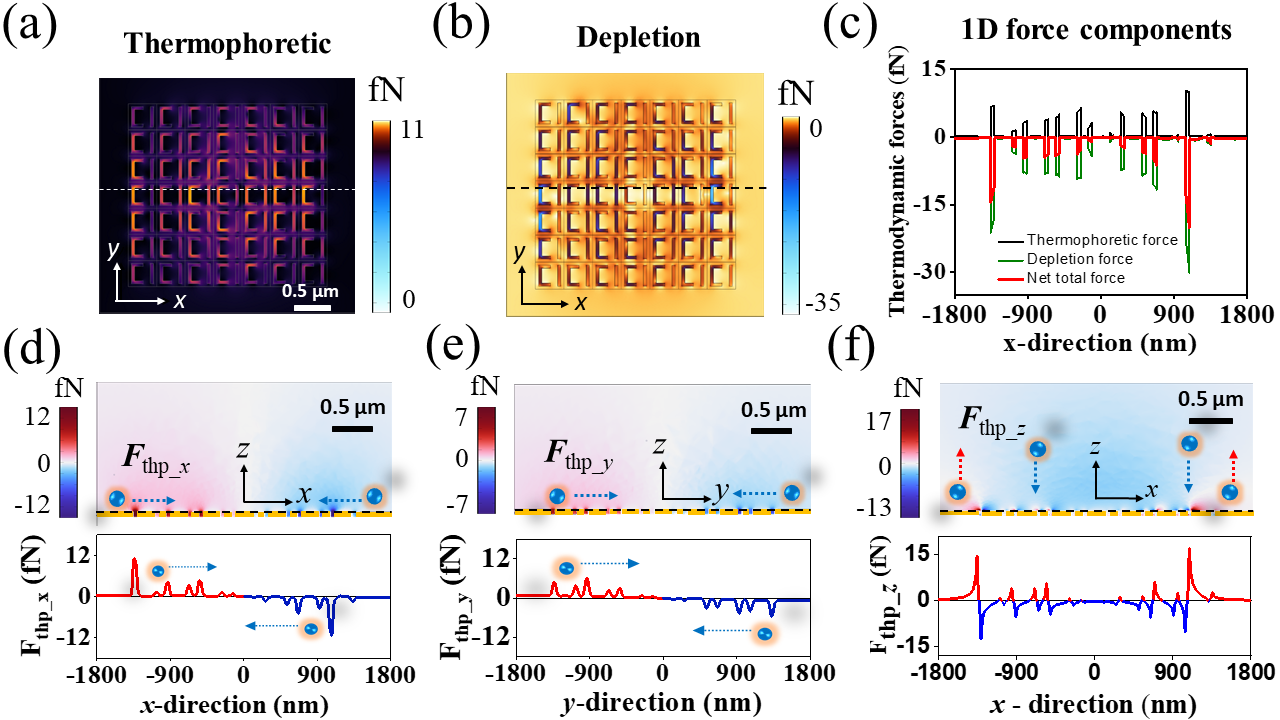}
\setlength\abovecaptionskip{0pt}
\caption{\label{Fig.2}\textbf{Thermodynamic forces exerted on an urease.} 2D map of (a) thermophoretic and (b) depletion forces in the \textit{xy}-plane at \textit{z} = \SI{7.4}{\nano\meter}. (c) 1D plots in (a) and (b) along the \textit{x}-direction at \textit{y} = \SI{110}{\nano\meter} (horizontal dashed line).
The black curve in (c) shows the positive thermophoretic (thermophobic motion), which pushes the urease away from its plasmonic trapping sites, while the green curve shows the negative depletion force (thermophilic motion) induced by the presence of the depletants in the solution, \textit{i.e.}, PEG$_{6000}$ molecules, which pulls the urease toward the ASR metamaterial surface. Negative values of the net total force,~\textit{i.e.}, the sum of thermophoretic and depletion force (red curve), \textbf{F$_{thp.}$} = \textbf{F$_{th.}$} + \textbf{F$_{dep.}$}, indicating the large contribution of the depletants to the trapping mechanisms. Figures (d), (e) and (f): (Top) 2D map of the total force components,~$\boldsymbol{F}_{{thp-x}}$, $\boldsymbol{F}_{{thp-y}}$, and $\boldsymbol{F}_{{thp-z}}$, in the \textit{zx}- and \textit{zy}-planes, and (bottom) their corresponding 1D line plots at \textit{z} = \SI{7.4}{\nano\meter} along the dashed black lines. 
Note that the \textit{x}- and \textit{y}-coordinates,~\textit{i.e.}, \textit{x} = -\SI{56}{\nano\meter} (along the \textit{y}-direction), and \textit{y} = \SI{110}{\nano\meter} (along the \textit{x}-direction) in (d), (e), and (f) were selected in order to observe the maximum peak values of the total forces exerted on the urease near the ASR surface. Both force components exhibit peak magnitudes at positions around \SI{1200}{\nano\meter} before decaying to zero at the center (x = \SI{0}{\nano\meter}). The incident Gaussian beam at $\lambda_t$ = \SI{937}{\nano\meter} is polarized parallel to the \textit{x}-direction. The $I_{trap}$ was fixed at 0.5 mW/\si{\micro\meter\squared}.}
\end{figure}

Using the calibrated temperature profile, the temperature gradient, $\nabla T$, is calculated for a trapping laser intensity, $I_{trap}$ = 0.5 \si{\milli\watt\per\square\micro\meter} (see Figures~S4 and SI-S1.2). Significant enhancements for temperature gradients within the ASR nano-aperture regions are observed when compared with those along the 50 nm gold thin film surface. A dominant temperature gradient of 195 $\si{\kelvin\per\micro\meter}$, which is 5 times higher than the temperature gradient \textit{i.e.}, at \textit{x} = 0, \textit{y} = -200 nm on the gold film surface, is observed at approximately \SI{1080}{\nano\meter} away from the focal point along the \textit{x}-direction (Figures~S5 and SI-S1.2). Moreover, the generated temperature gradient along the gold ASR metamaterial/liquid interface will induce quasi-slip flow within the liquid boundary layer. This is because the liquid interacts differently with the gold surface at different temperatures, resulting in an excess enthalpy that drives a flow along the gold/water interface, known as thermo-osmotic flow~\cite{Bregulla}. For thermo-osmotic flow field simulation, we combine the heat transfer (HT) module with the laminar flow (LF) module (see Figure S6 and SI-S1.3).

Considering only a dilute urease solution in water, the ureases experience a thermophoretic drift in the temperature gradient that drives them from the hotter region toward the colder region. The thermophoretic drift velocity, \textbf{u}(\textbf{r}, \textit{t}), of the urease is described by the equation~\cite{Wurger_2010}, $\textbf{u}(\textbf{r},\textit{t}) $ = - $D_{T} \nabla T$, where $D_{T}$ indicating the urease thermophoretic mobility. Although the thermophoretic motion is force-free, the same drift speed could be obtained when dragging the molecule with an external force that is $\boldsymbol{F}=\eta \textbf{u}$. For better comparison with other forces we therefore report these equivalent force values.

When additional crowding molecules, such as PEG$_{6000}$, are added to the dilute urease solution, both PEG$_{6000}$ and urease molecules will undergo thermo-osmotic flows and thermophoretic drifts comparable to colloidal systems \cite{Quinn2025}. These thermo-osmotic flows will generate convection-like flow fields near the hot spot, although there is no convection present due to the sample geometry. The temperature gradient additionally induces thermophoretic drifts on all molecules in the solvent. 
In the steady state, the molecular concentration gradients can be estimated by~\cite{Sano} $\nabla c$ = -c$S_{T} \nabla T$, where \textit{c} is the concentration ($\si{\mol\per\litre}$), $S_{T}$ = $D_{T}/D$ is the Soret coefficient, and \textit{D} represents the diffusion coefficient of the urease or the PEG$_{6000}$ molecules. To the best of our knowledge, there has been no direct experimental determination of the Soret coefficient, $S_{T,U}$ for urease reported in the literature until now. It has been suggested that thermodiffusion behaviors strongly depend on hydrophobic interactions of proteins in solutions~\cite{Iacopini_2003}. Thus, the protein migration directions were associated with the exposure of hydrophobic groups to the solvent~\cite{Iacopini_2003}. Typically, $S_{T}$ of proteins is positive and insignificant change at temperatures higher than the ambient temperature~\cite{Iacopini_2003}. For this reason, we assume the Soret coefficient, $S_{T,U}$ = 0.02 $\si{\per\kelvin}$, for a \SI{14.8}{\nano\meter} urease, which is very close to the values of $S_{T}$ = 0.01~-~0.02 $\si{\per\kelvin}$ for \SI{3.6}{\nano\meter} proteins of T4 Lysozyme catalytic enzyme and their mutant variants for temperature increments larger than $\Delta T=\SI{25}{\kelvin}$~\cite{Putnam,Pu}. The PEG$_{6000}$ molecule Soret coefficient is set at $S_{T,PEG}$ = 0.056 $\si{\per\kelvin}$ (see SI-S1.3 and S1.4). Figure~\ref{Fig.2}(a) shows a 2D spatial distribution of the thermophoretic force acting on a single urease over the surface of the illuminated 7~$\times$~7 ASR metamolecules for an $I_{trap}$ = 0.5 $\, \si{\milli\watt\per\micro\meter\squared}$. Positive values of the thermophoretic force indicate repulsive forces pushing the urease away from the ASR metamolecule surfaces.

Under dilute urease conditions, where the enzyme concentration is lower than that of PEG$_{6000}$, the PEG concentration is highly depleted from the heated region at the center focal point of the incident Gaussian beam (see Figures S7 (see SI-S1.3)). The forming concentration gradient of the PEG$_{6000}$ molecules will induce an osmotic pressure to cause a diffusiophoretic drift (a temperature induced depletion drift) on the urease, directing them toward the region of lower PEG crowder density (see SI-S1.4). Figure~\ref{Fig.2}(b) shows a 2D spatial distribution of the depletion force exerted on the urease molecules. Negative values of the depletion force imply attractive interactions that pull the urease toward the ASR metamolecules. As a result, by introducing PEG$_{6000}$ into the solution, the overall drift of ureases can be modulated from thermophobic behavior (driven by a positive thermophoretic force) to thermophilic behavior (driven by a negative depletion force), as shown in Figure~\ref{Fig.2}(c).

As a consequence, the total force resulting from the thermophoretic and depletion drift vector fields provides multiple stable trapping sites within the sample plane. Figures~\ref{Fig.2}(d) and (e) (top) show the 2D spatial distributions of the total force field components, $\boldsymbol{F}_{{thp-x}}$, and $\boldsymbol{F}_{{thp-y}}$, which are responsible for the movement of the urease along the ASR surface, while the bottom curves indicating their corresponding 1D line plots taken horizontally at \textit{z} = \SI{7.4}{\nano\meter} (similar to the urease's hydrodynamic radius). The vertical, $\boldsymbol{F}_{{thp-z}}$, component in Figure~\ref{Fig.2}(f) is induced by the longitudinal temperature gradient part (see Figure S4(c) and SI-S1.2) along the \textit{z}-direction of the ASR nano-apertures. Significant enhancements for all simulated thermal forces are observed within the nano-aperture regions. A negative total, $\boldsymbol{F}_{{thp-z}}$, force component (blue curve in Figure~\ref{Fig.2}(f)) indicates that the urease tends to vertically pull towards the thermal hot-spot locations of the ASR nano-apertures, while positive values (red curve in Figure~\ref{Fig.2}(f)) present repulsive forces, which push the urease away from the ASR surface. Apparently, the longitudinal component, ${F_{thp-z}}$, has larger magnitude than the radial force components, ${F_{thp-x}}$ or ${F_{thp-y}}$ \textit{i.e.}, maximum of \SI{17}{\femto\newton} for the ${F_{thp-z}}$ \textit{vs} the most dominant of \SI{10}{\femto\newton} for the ${F_{thp-x}}$ part. This implies that the urease will first propel along the ASR metamolecule surface driven by $\boldsymbol{F}_{{thp-x}}$ or $\boldsymbol{F}_{{thp-y}}$, and hence, trap at the ASR thermal hot-spots due to the presence of the more pronounced, $\boldsymbol{F}_{{thp-z}}$, force component.

\noindent \textbf{Off-center plasmonic thermodynamic trapping}

\begin{figure}[h]
\centering
\includegraphics[width=0.90\textwidth]{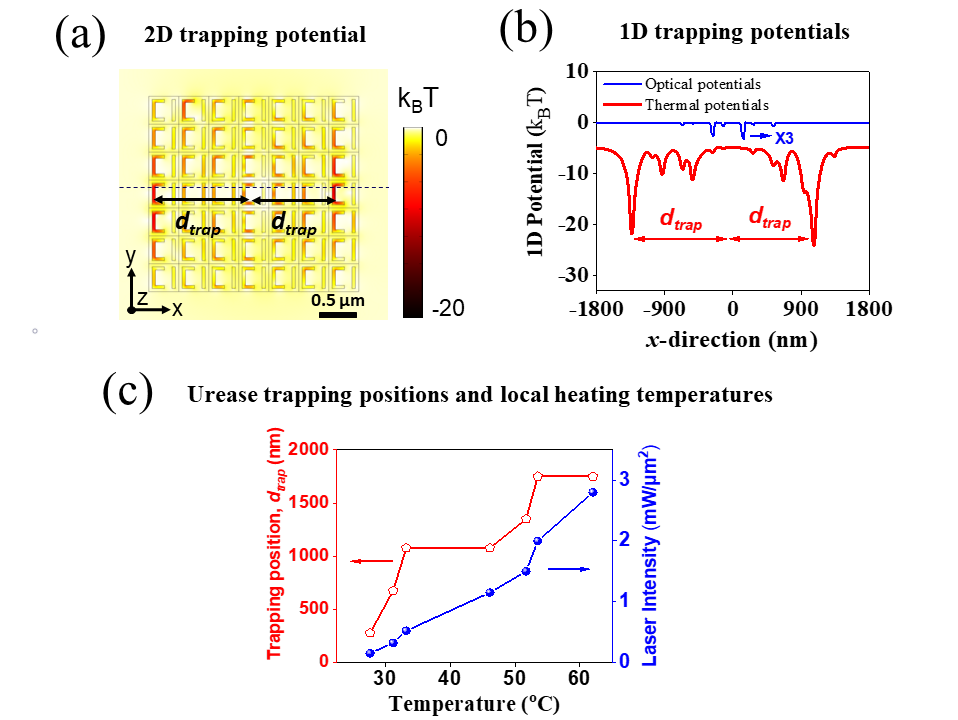}
\setlength\abovecaptionskip{10pt}
\caption{\label{Fig.3}\textbf{Trapping positions and local heating temperatures acting on trapped enzymes}.
 (a) Top view of the 2D trapping potential on a urease with 
 a diameter of \SI{14.8}{\nano\meter} in PEG$_{6000}$/water solution (see SI-S2.1). The trapping potential is calculated by taking the sum of the longitudinal trapping thermodynamic force component, including the thermophoretic, and the nonequilibrium depletion forces at the ASR metamolecule's surface. (b) 1D trapping potential line plots of trapped urease, extracted at \textit{y} = \SI{110}{\nano\meter} along the \textit{x}-direction (horizontal dashed black line in (a)). The blue line represents the optical trapping potential, localized near the metamaterial's center (see Figure S9 and SI-S1.4), while the red line shows the net thermal potential well, which extends far from the center. $I_{trap}$ is fixed at 0.5 mW/\si{\micro\meter\squared}. (c) Trapping position, $\textit{d}_{trap}$ (red points), indicates the distance away from the center focal point at \SI{0}{\nano\meter} of the incident Gaussian beam to the minimum trapping potential depth position, and its corresponding $I_{trap}$ (blue points) as a function of temperature heating on the trapped enzyme. The solid lines are a guide for the eye. Note that the incident Gaussian beam at $\lambda_{t}$ = \SI{937}{\nano\meter} is polarized parallel to the $\textit{x}$-direction.
}
\end{figure}
In Figure~\ref{Fig.3}(a), we show the 2D trapping potentials of a trapped urease molecule with a hydrodynamic radius of \SI{7.4}{\nano\meter} suspended in an aqueous PEG$_{6000}$/water (see SI-S2.1). The thermodynamic potential in the \textit{xy}-plane was generated by numerically integrating the longitudinal trapping force, $\boldsymbol{F}_{{thp-z}}$, component across the metamaterial surface. The optical potential depth is more dominant along the \textit{y}- than the \textit{x}-direction (see Figure S9 and SI-S1.5), while the thermal depths behave in the opposite way. Moreover, Figure~\ref{Fig.3}(b) illustrates that the optical potential wells exhibit shallower depths as radial distances, $x$, increase from the nanostructure's center. Conversely, the thermal potential wells deepen with increasing radial distances, $d_{trap}$. It is noteworthy that the optical potentials, with the deepest minima depths of 1.5 k$_{B}$T in the \textit{x}-direction and 2.5 k$_{B}$T in the \textit{y}-direction, are relatively small compared to the thermodynamic potential wells, which reach depths of 24.5 k$_{B}$T for $d_{trap}$ = \SI{1080}{\nano\meter} under resonant illumination, $\lambda_{t}$ = \SI{937}{\nano\meter}, and $I_{trap}$ =~0.5 \si{\milli\watt\per\micro\meter\squared}.  Yet, this optical force alone is insufficient to trap urease molecules at low trapping laser intensities. The total thermodynamic force leads to trapping potential depths greater than 10 k$_{B}$T, which is enough to oppose the Brownian motion of the urease suspended in
solution.
At low trapping laser intensities, \textit{i.e.}, $I_{trap}$ = 0.35 - 1 \si{\milli\watt\per\micro\meter\squared}, the maximum trapping potential
depths were calculated in the range of 15 k$_{B}$T to 24.5 k$_{B}$T, reaching its optimal stable trapping at  $I_{trap}$ = 0.5 \si{\milli\watt\per\square\micro\meter}; therefore, urease seems to be stably trapped in off-center positions.
It is important to note that these observations are based on the assumption that urease has a value of $S_{T,U}$ in the range of 0.01 - 0.02 \si{\per\kelvin} (see SI-S1.4). To further validate this hypothesis, we theoretically analyzed the magnitude of the thermophoretic force for $S_{T,U}$ between 0.01 - 0.05 \si{\per\kelvin} and found that, when the urease Soret coefficient exceeds 0.05 \si{\per\kelvin}, the thermophoretic force becomes dominant enough to push the urease away from the trapping sites, preventing trapping (see Figure S8 and SI-S1.4). Based on the theoretical results and experimental observations of the trapped urease at various trapping laser intensities and trapping positions (Figure~\ref{Fig.3}(c)), we can conclude that the Soret coefficient, $S_{T,U}$, for urease would likely lie within the narrow range of 0.01 - 0.02 \si{\per\kelvin}, confirming our initial hypothesis.

Finally, Figure~\ref{Fig.3}(c) shows the trapping position of the urease (red line), $d_{trap}$, and the trapping laser intensity, $I_{trap}$ (blue line), as a function of the absolute temperature, T$_{trap}$(\textbf{r}). Here, T$_{trap}$(\textbf{r}) values are identified at the $d_{trap}$ position of the trapped urease. As we increase the $I_{trap}$, there are three key processes that determine the trapping position. First, the osmotic-flow field (see Figure S6 and SI-S1.3) circulates the urease toward the ASR metamaterial plasmonic hotspot surface. Then, the optical force tends to trap the urease near the center of ASR metamolecules, while the accompanying thermal gradient generates a thermophoretic force that pushes the urease enzymes away from the hot center of the ASR metamaterial. Finally, the diffusiophoretic (depletion) force, generated by the presence of the PEG$_{6000}$ crowders in the solution, will assist in trapping the urease by creating an additional attractive effect. The enzyme ultimately settles at an intermediate position, $d_{trap}$, where these forces balance - far enough from the center to avoid strong thermophoretic pushing but close enough to benefit from the osmotic-fluid flow drift, the weaker optical trapping force, and the more pronounced diffusiophoretic drift (depletion) assistance (see SI-S1.3, S1.4, and S1.5). This four-way competition may explain why we observe different stable trapping locations that varying with laser intensity, as shown in Figures~\ref{Fig.3}(b) and (c).

\noindent \textbf{Real-time detection of trapped target particles using an avalanche photodiode (APD)}\

Figure~\ref{Fig.4}(a) shows the time evolution of the transmitted signal as measured on the APD through the ASR metamolecules in the presence of urease solution. The abrupt jump in the APD signal at 21~s and the change in the noise level right after the jump are signatures of biomolecular trapping by the metamaterial. Additionally, the cycle of trapping and then releasing the enzyme was repeated by blocking and unblocking the trapping laser beam.

\begin{figure}
    \centering \captionsetup{justification=raggedright,singlelinecheck=false}
        \includegraphics[trim={0cm 0.5cm 0.5cm 0.4cm},clip,width=0.85\textwidth]{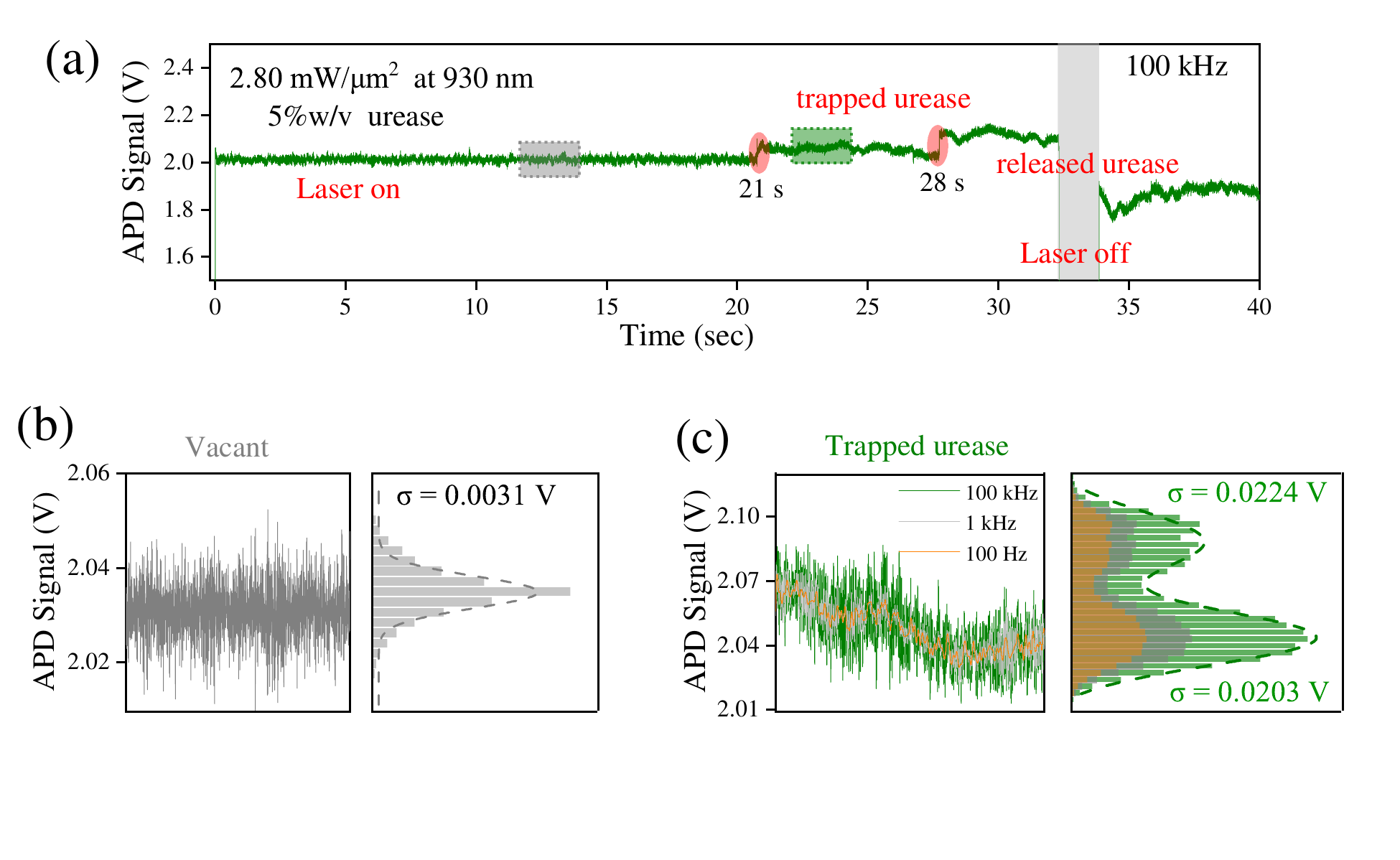}
         \includegraphics[trim={0cm 0.5cm 0.5cm 0.4cm},clip,width=0.85\textwidth]{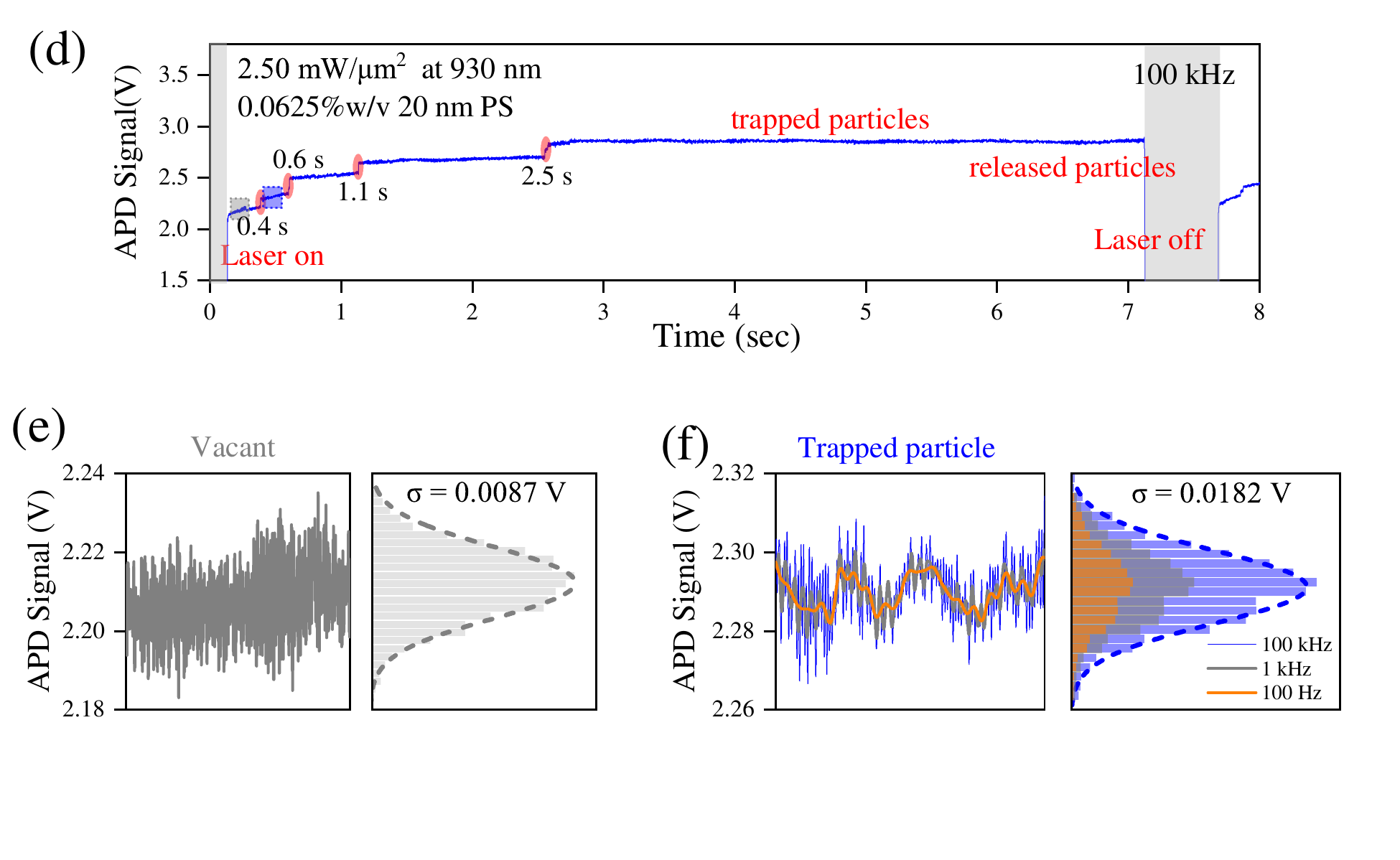}
\setlength\abovecaptionskip{-20pt}
 \captionsetup{justification=justified}
 \caption{\textbf{Optical trapping of the target objects in action}. (a) Time trace of optical transmission through the ASR metamaterial, showing the optical trapping of a single urease as a sudden discrete jump in the APD signal at \SI{21}{s}. A few seconds later another jump indicates a second trapping event (at \SI{28}{s}). A trapping laser wavelength of \SI{930}{nm} and trapping laser intensity, $I_{trap}$ at the sample plane of \SI{2.80}{\milli\watt\per\square\micro\meter} were used. Transmission signal fluctuations and the Gaussian fit (dashed lines) to the corresponding histogram of the transmitted signal fluctuations in case of (b) vacant (gray box in (a)) and (c) trapped (green box in (a)) states of the urease. (d) Time trace of optical transmission through the ASR metamaterial, showing the optical trapping of a \SI{20}{nm} PS particle at shorter times (first trapping jump at \SI{0.4}{s}, second at \SI{0.6}{s}, third at \SI{1.1}{s} and forth at \SI{2.5}{s}). A $I_{trap}$ at the sample plane of \SI{2.50}{\milli\watt\per\square\micro\meter} was used. Transmission signal fluctuations and the Gaussian fit (dashed lines) to the corresponding histogram of the (e) vacant (gray box in (d)) and (f) trapped (blue box in (d)) states of the PS nanoparticle. The width of the Gaussian distribution, $\sigma$, shows the root-mean-square (RMS) variation of the transmitted signal.}
        \label{Fig.4}
\end{figure}

Figures~\ref{Fig.4}(b) and (c) show the transmitted signal fluctuations with corresponding histograms for the two different states,~\textit{i.e.}, vacant and trapped, as determined from raw data collected at \SI{100}{\kilo\hertz}. The width of the Gaussian distribution for the vacant case displays a small root-mean-square (RMS) deviation, $\sigma$~=~\SI{0.0031}{\volt}, while the RMS deviation for the trapped event, $\sigma$~=~\SI{0.0203}{\volt}, for the second peak (or $\sigma$~=~\SI{0.0224}{\volt} for the first peak), is larger. For comparison, we show a sequence of trapping and releasing events of a \SI{20}{\nano\meter} PS particle in an array of ASR metamolecules in Figure~\ref{Fig.4}(d). As we illuminated 3~$\times$~3 ASR metamolecules, we excited eighteen plasmonic hotspots, which can trap up to eighteen nanoparticles simultaneously~\cite{DomnaAPL,Bouloumis}. Therefore, we assume that each jump in the APD signal corresponded to a single trapping event. Figures~\ref{Fig.4}(e) and (f) show fluctuations in the APD signal with corresponding histograms as well as the Gaussian fit for the trapped nanoparticle and the vacant positions. Similarly to the case for urease, we observe that the APD signal for trapped PS nanoparticles has a larger RMS deviation than that observed for no trapping event. Comparing Figures~\ref{Fig.4}(c) and (f), we notice that, in the case of protein trapping, the signal fluctuations are larger than for PS particle trapping. We assume that the hydrodynamic motion and conformation changes of the protein in the ASR metamolecule may increase fluctuations in the transmitted intensity of the trapping laser~\cite{Young}.

\newpage
\noindent \textbf{Statistical thermodynamics of trapped objects}\

In addition, we calculated the probability density function (PDF) of the APD signal through the metamaterial both before and after trapping urease with a concentration of \SI{5}{\wpv}, and PS nanoparticles for various laser intensities (Figure~\ref{Fig.5}). In this case, the PDF refers to the likelihood of finding the trapped particle in a specific arrangement or exhibiting a particular characteristic~\cite{Madan}. We observe that trapping urease leads to a PDF with a wider amplitude distribution with two or more peaks compared to the vacant signal (Figures~\ref{Fig.5}(b)). This might indicate heterogeneity in enzyme activity, conformational states, or interactions with the environment~\cite{Bustamante}. However, the PDF calculated from the PS nanoparticle trapping signal shows a relatively narrow peak compared to the vacant signal (Figures~\ref{Fig.5}(d)). This observation reveals that the PS nanoparticles behave more uniformly or consistently when trapped.

\begin{figure}[H]
     \centering \captionsetup{justification=raggedright,singlelinecheck=false}
        \includegraphics[trim={0cm 0cm 0cm 0cm},clip,width=0.6\textwidth]{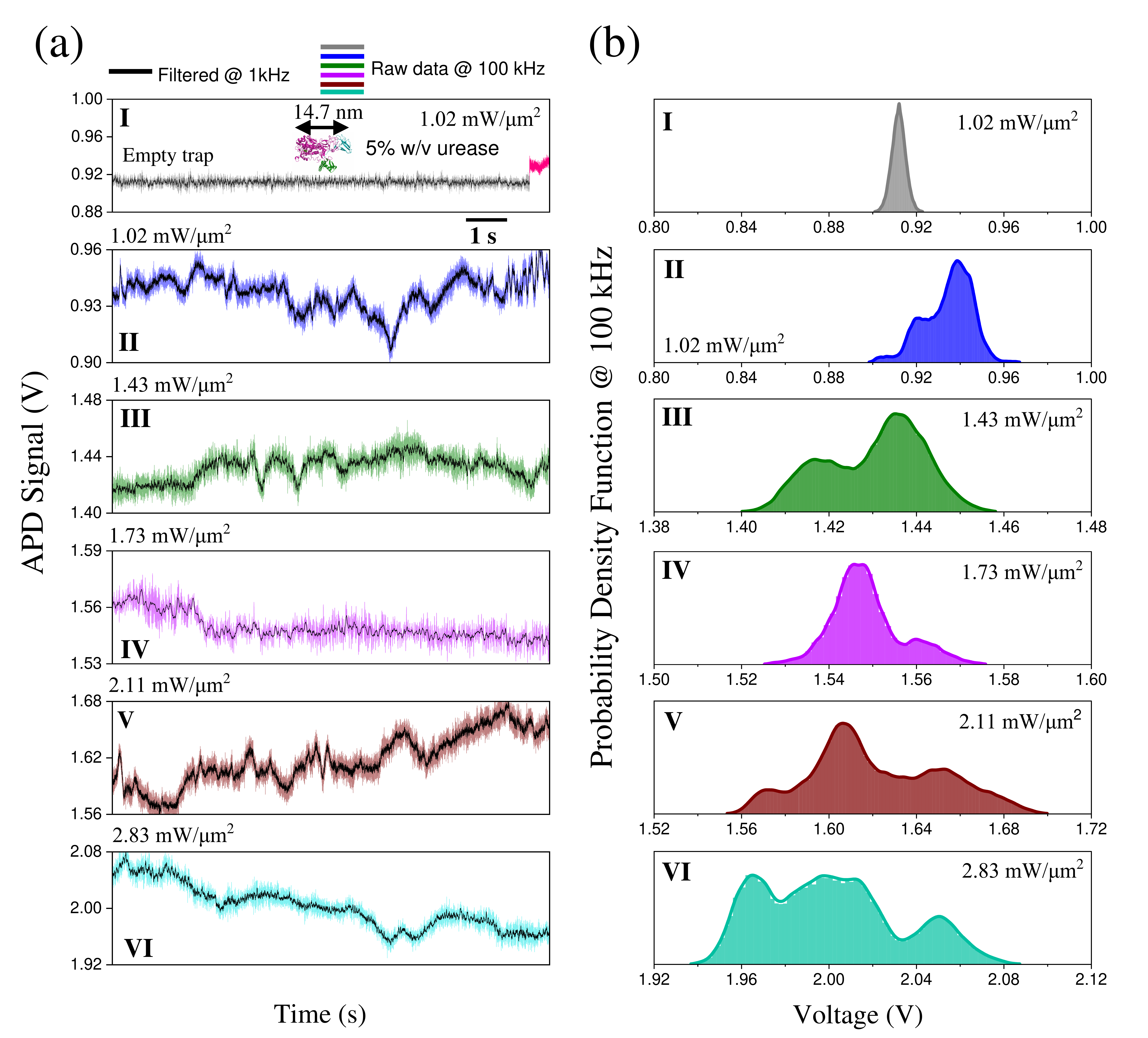}
     \includegraphics[trim={0cm 0cm 0cm 0cm},clip,width=0.6\textwidth]{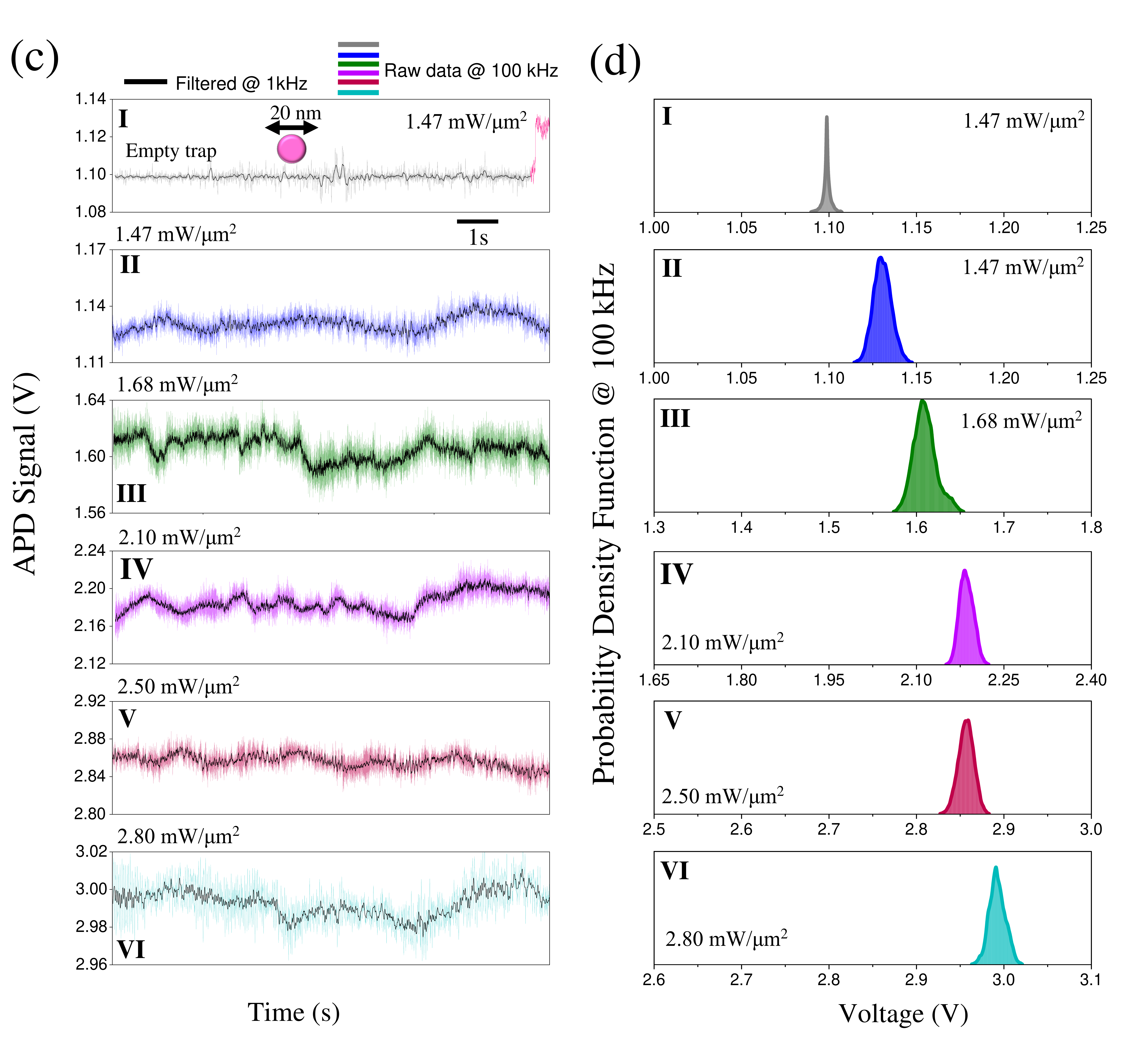}
\setlength\abovecaptionskip{0pt}
 \captionsetup{justification=justified}
        \caption{\textbf{Single urease trapping and probability density function analysis}. (a) Segments of the time trace of the transmission signal through the ASR metamaterial: (I) without urease in the trapping position at a $I_{trap}$ of \SI{1.02}{\milli\watt\per\square\micro\meter}, and (II-VI) with urease in the trapping position at various trapping laser intensities. A urease concentration of \SI{5}{\wpv} was used.(b) Probability density function (PDF) of the APD signal calculated from (I) traces before and after trapping (II-VI) for \SI{7.4}{\nano\meter} urease. (c) Segments of the transmission signal through the ASR metamaterial (I) before trapping at \SI{1.47}{\milli\watt\per\square\micro\meter}, and (II-VI) after trapping of a single \SI{20}{\nano\meter} PS particles at various trapping laser intensities. (d) PDF of the APD signal calculated from time trace signals of (c). The solid line in all graphs fits the kernel density distribution. The PDF was calculated from the raw data collected, without using low-pass filters (\SI{100}{\kilo\hertz}) (see SI-S4). A trapping laser wavelength of \SI{930}{\nano\meter} and a time duration of \SI{12}{\second} were used to calculate the PDFs.}
        \label{Fig.5}
\end{figure}
To further investigate this observation, we calculated the PDFs of the changes in the APD signal through the metamaterial for successive trapping events for both urease and PS nanoparticles. For this purpose, we defined and calculated the relative transmitted APD signal as $(T_{i}-T_{i-1})/{T_{i-1}}$, where \textit{$T_{i}$} is the transmitted signal for each trapping event, \textit{$T_{i-1}$} is the transmitted signal one trapping step/jump before and \textit{i} is the number of trapping events ($i = 1, 2, 3, 4...$ trapping steps). This method minimizes effects caused by previously trapped particles, the metamaterial structure, and instrumentation noise. Note that by changing the trapping laser power and the trapping duration, we observed single or multiple trapping events.
In Figure~\ref{Fig.6}(a), we present violin plots for both types of particles at the lowest $I_{trap}$ for which we observed trapping.
The shape of the violin plot represents the PDF, which is used to calculate the free energy (see SI-S5) and to understand the behavior of the trapped particle in various trapping conditions.
By comparing the plots in Figure~\ref{Fig.6}(a), we observe that the median value of the trapping data in the urease violin plot is 0.54, while for PS particles it is 0.52 as a result of normalization (white dot in Figure~\ref{Fig.6}(a)). We also notice that the PDF of urease exhibits two distinct peaks, indicating a bimodal-like distribution, while the PDF for PS nanoparticles is more homogeneous, showing a single, symmetrical peak characteristic of a normal distribution.
\begin{figure}[]
    \centering \captionsetup{justification=raggedright,singlelinecheck=false}
      \includegraphics[trim={0.5cm 0.5cm 0.3cm 0.3cm},clip,width=0.75\textwidth]{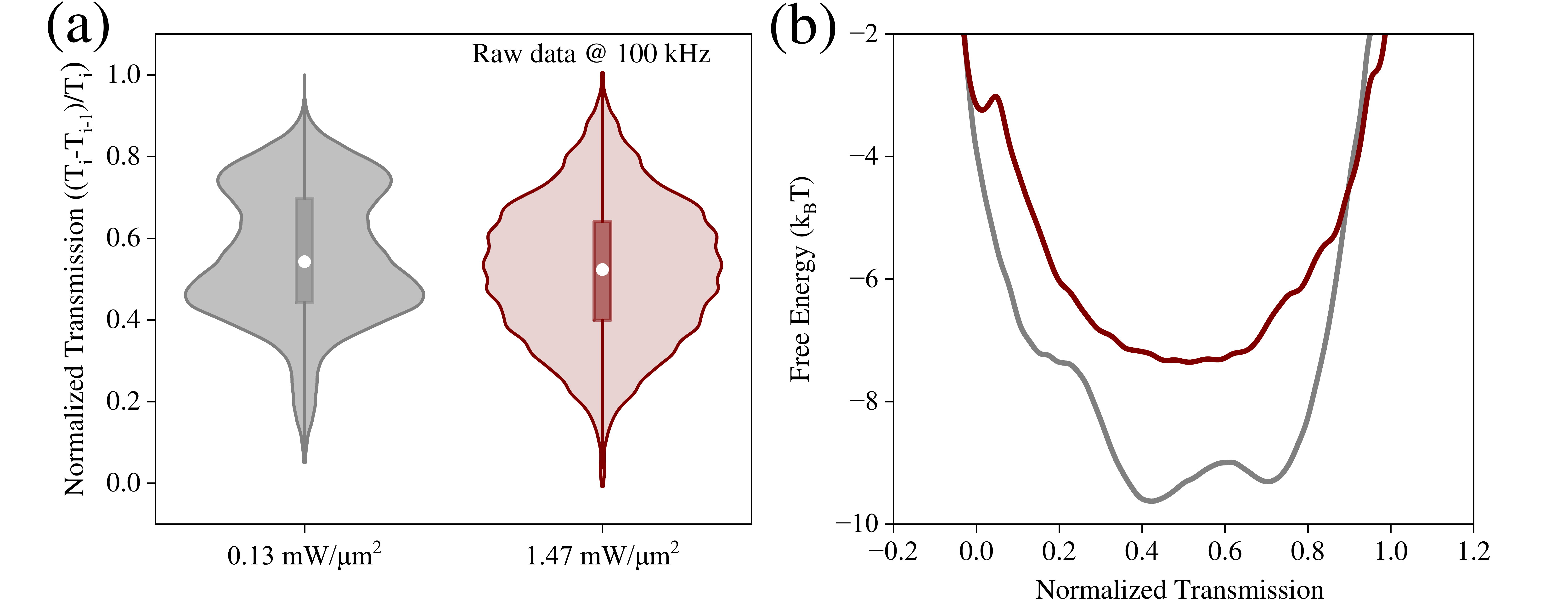}
      \includegraphics[trim={0.5cm 0.5cm 0.3cm 0.5cm},clip,width=0.75\textwidth]{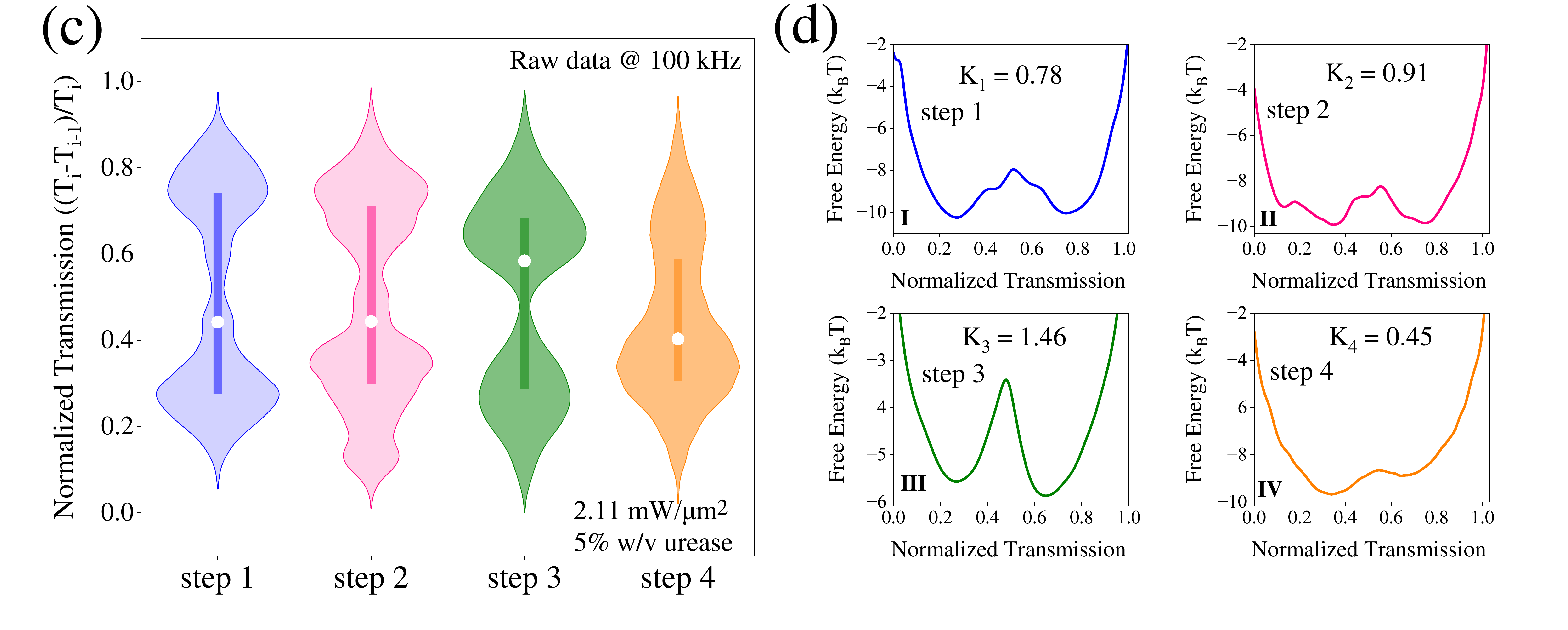}
      \includegraphics[trim={0.5cm 0.5cm 0.3cm 0.1cm},clip,width=0.75\textwidth, center]{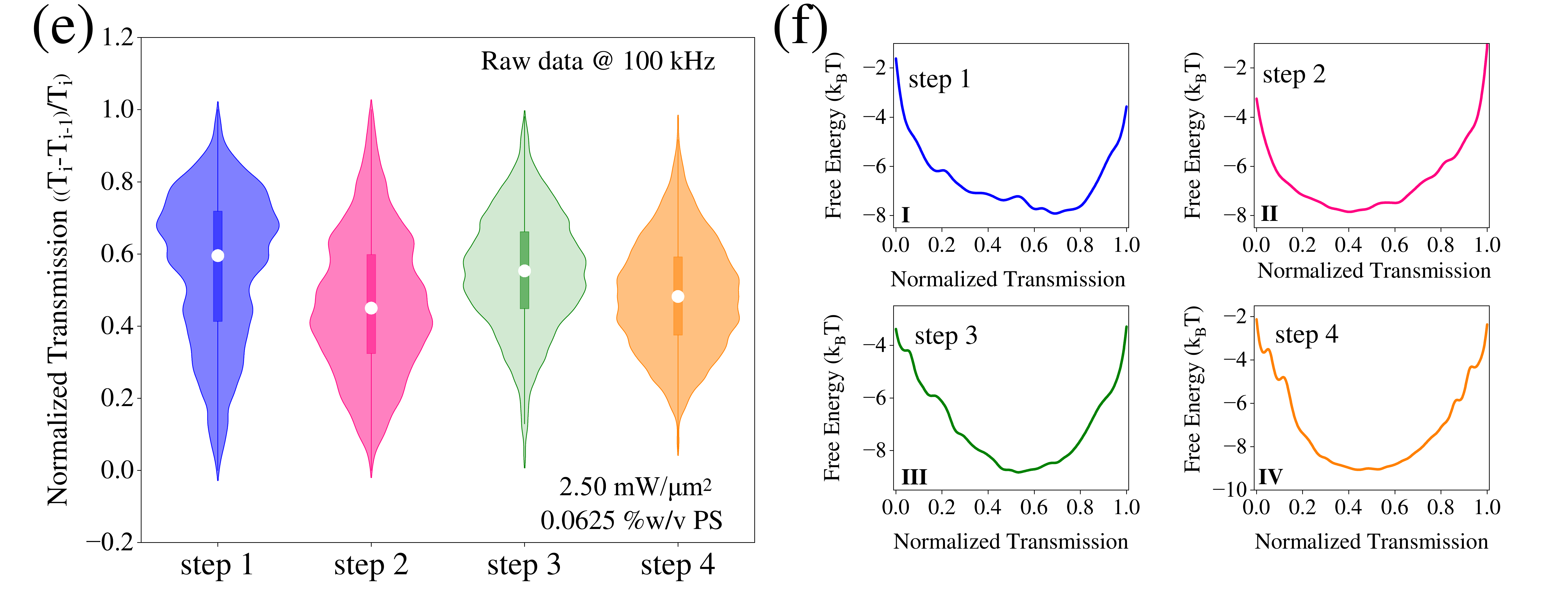}
  \setlength\abovecaptionskip{0pt}
 \captionsetup{justification=justified}
\caption{\textbf{Statistical analysis and free energy profiles}. (a) Violin plots for the first trapping event for a urease molecule with a concentration of \SI{2.5}{\wpv} (gray plot) and for a \SI{20}{\nano\meter} PS particle with a concentration of \SI{0.0625}{\wpv} (light wine plot). For urease trapping, the \textit{I$_{trap}$} was \SI{0.13}{\milli\watt\per\square\micro\meter} (T$\sim$\SI{27}{\celsius} and $d_{trap}$ = \SI{265}{\nano\meter}) while for PS nanoparticles it was \SI{1.4}{\milli\watt\per\square\micro\meter} (T$\sim$\SI{50}{\celsius} and $d_{trap}$ = \SI{1246}{\nano\meter}). In both violin plots, the white dots indicate the median value, which is 0.54 for urease and 0.52 for PS particles. (b) The free energy profile of both particle types under the conditions described in (a). (c) Violin plots and (d) the free energy profile for successive trapping events of urease molecules that are noted as step 1,2,3 and 4 at a \textit{I$_{trap}$} of \SI{2.11}{\milli\watt\per\square\micro\meter} (T$\sim$\SI{57}{\celsius} and $d_{trap}$ = \SI{1600}{\nano\meter}) and a concentration of \SI{5.0}{\wpv}. The equilibrium constant is noted for each trapping step as calculated by the free energy profile (see SI-S5). The media values for the step 1,2,3, and 4 are 0.44, 0.44, 0.58, and 0.40 respectively. (e) Violin plots and (f) the free energy for multiple trapping events that are noted as step 1, 2, 3, and 4 of \SI{20}{\nano\meter} PS particles at a \textit{I$_{trap}$} of \SI{2.50}{\milli\watt\per\square\micro\meter} (T$\sim$\SI{60}{\celsius} and $d_{trap}$ = \SI{1690}{\nano\meter}) and a concentration of \SI{0.0625}{\wpv}. The media values for the step 1,2,3, and 4 are 0.59, 0.45, 0.55, and 0.48, respectively. The violin plots and the free energy were calculated from the raw data collected at \SI{100}{\kilo\hertz}. The wavelength of the trapping laser was \SI{930}{\nano\meter}.}
\label{Fig.6}
\end{figure}
Since we did not observe multiple urease trapping events for every $I_{trap}$ applied to illuminate the metamaterial, whereas we did for the PS particles, we analyzed data where up to four trapping events were observed for both types of particles at slightly different \textit{I$_{trap}$}. These successive trapping events are labeled as step 1, 2, 3, and 4 in Figures~\ref{Fig.6}(c) and (e).
Likewise, in the case of enzyme trapping at high trapping laser intensities where successive trapping events occur,
two peaks appear in the violin plots (Figure~\ref{Fig.6}(c)).  The median values for the trapping events at step 1 to 4 are 0.44, 0.44, 0.58, and 0.40, respectively. These observations may suggest that the first, second, and fourth trapped urease, with slightly different median values, may exhibit comparable behavior in their trapping position. However, the increase in the median value at step 3 (0.58) compared to the other steps suggests that the urease could be experiencing a shift in its dynamics at that particular stage. This deviation could arise from factors such as changes in the molecule's conformation or interactions with the surrounding environment during trapping.

Figure~\ref{Fig.6}(d) shows the free energy profiles for each trapping event in the case of urease with a concentration of \SI{5}{\wpv} and \textit{I$_{trap}$} at \SI{2.11}{\milli\watt\per\square\micro\meter} (see SI-S5). We notice that the free energy landscape for the four successive trapping events reveals distinct profiles, each characterized by specific energy minima and barriers. This may suggest that the enzyme's behavior is highly sensitive to its local environment, including the specific ASR hotspot where it is trapped. For example, the first trapping event (Figure~\ref{Fig.6}(d)(I)) exhibits a free energy profile with two minima of slightly different magnitudes, indicating that the enzyme may adopt two distinct conformational states during this event. However, the second trapping event (Figure~\ref{Fig.6}(d)(II)) shows a more complex energy profile, with a broader distribution of energy minima, suggesting that the enzyme may explore a wider range of conformational states. The third trapping event (Figure~\ref{Fig.6}(d)(III)) reveals a large energy barrier compared to steps 1 and 2, indicating a significant change in the enzyme's conformation or orientation within the plasmonic hotspot. Finally, the fourth trapping event (Figure~\ref{Fig.6}(d)(IV)) returns to a profile similar to step 1 but with a different equilibrium constant (see Discussion), suggesting that the enzyme has settled into a different stable state.

Similarly, Figure~\ref{Fig.6}(e) shows the violin plots of four PS nanoparticle trapping events. Due to polystyrene's homogeneous nature, all violin plots show one peak, indicating that the data may have similar characteristics.  This uniformity arises from the rigid structure and the lack of internal degrees of freedom in contrast to what was observed for ureases. As a result, the free energy profiles for the nanoparticles are due to their translational motion in the trapping potential.

\section{Discussion}

Proteins, such as enzymes, are held together by various weak interactions, including ionic bonds, van der Waals forces, and hydrophobic interactions. As the temperature rises, the thermal energy can overcome these weak interactions, leading to changes in protein conformation. Since urease is sensitive to changes in temperature~\cite{Feder}, its structural conformation can be modulated by the local temperature increase on the metamaterial.
We calculated the thermodynamic equilibrium constant, \textit{K}, of 0.78, 0.91, 1.46, 0.45 for urease particles that have been trapped successively and noted as steps 1, 2, 3, and 4, respectively (Figure
~\ref{Fig.6}(d)), using the free energy profiles (see SI-S5). A large \textit{K} magnitude indicates a favored conformation stage. In this case, step 3 provides the largest \textit{K} values and may reflect a large conformation of the enzyme during this trapping event. Based on this observation, we calculate the absolute value of the Gibbs energy difference for step 3 equal to $\Delta$G =  \SI{0.25}{\kilo\calorie\per\mole} (the absolute temperature corresponding to the \textit{I$_{trap}$} =~\SI{2.11}{\milli\watt\per\square\micro\meter} is \SI{330}{\kelvin} (Figure~\ref{Fig.3}(c)). However, the literature reports a much larger Gibbs energy change magnitude for urease unfolding under chemical denaturation conditions, with \SI{19.0}{\kilo\calorie\per\mole} for the dissociation of dimers into monomers~\cite{OMAR}. While the literature value reflects the energy required to fully denature urease obtained by fluorescence microscopy measurements, our calculation represents a smaller energy change associated with conformational transitions within the trapped enzyme, emphasizing the enzyme's sensitivity to its local environment. Notably, fluorescence microscopy typically requires labeling, which may perturb the system, whereas the proposed metamaterial tweezers enable label-free detection, providing a more direct measurement of enzyme dynamics.

It has been reported that the conformation of urease remains stable below \SI{40}{\celsius}, corresponding to the native state~\cite{OMAR1994}. Then, a dissociation of the hexamer into folded dissociation products is caused between \SI{50}{\celsius}~-~\SI{65}{\celsius}~\cite{OMAR1994}. A more significant structural transition has been observed upon further heating at about \SI{72}{\celsius}, indicative of a change in protein structure leading to further exposure of tryptophan to water, revealing that the monomers are unfolded~\cite{OMAR}. This denaturation process leads to complete thermal inactivation of the urease above \SI{80}{\celsius}~\cite{Grancic}. The observed energy landscapes (Figure~\ref{Fig.6}(d)), recorded at \SI{57}{\celsius} (\SI{2.11}{\milli\watt\per\square\micro\meter} in Figure~\ref{Fig.3}(c)), align with previous reports~\cite{OMAR1994,OMAR}, suggesting that the urease undergoes dissociation into folded subunits at this temperature. It should be noted that the optical trapping data presented in this work were obtained using a solution containing \SI{5}{\wpv} urease in \SI{4}{\wpv} PEG$_{6000}$/water solution. This concentration of urease compared to PEG$_{6000}$ molecules may lead to PEG$_{6000}$ trapping due to the possible crowding effect induced by urease. Since the melting point range for the PEG$_{6000}$ is \SI{60}{\celsius} - \SI{63}{\celsius} (SDS, CAS No:25322-68-3(81260), Sigma-Aldrich), and the $I_{trap}$ for which we observed trapping events corresponds to a temperature range of \SI{57}{\celsius} - \SI{62}{\celsius}, we may assume that we avoid PEG$_{6000}$ trapping. In addition, PEG$_{6000}$ can improve the catalytic activity of urease as the temperature increases up to \SI{60}{\celsius} (see SI-S2). Notably, prior work has demonstrated that an optimal 1:1 molar ratio of urease to PEG derivatives maximizes enzymatic stability~\cite{Vardar}. Our experimental conditions (\SI{5}{\wpv} urease and \SI{4}{\wpv} PEG$_{6000}$) were carefully selected to approximate this stoichiometric balance for optimal enzyme stabilization, and to maintain a sufficient PEG$_{6000}$ concentration to drive urease accumulation at plasmonic hotspots through molecular crowding effects. For temperatures larger than \SI{70}{\celsius}, thermal inactivation of urease was observed to be irreversible due to aggregation formation~\cite{OMAR1994, OMAR,Grancic}. However, the range of $I_{trap}$ that we have used in this work provides a theoretical threshold of \SI{62}{\celsius} (Figure~\ref{Fig.3}(c)), meaning that the system remains in physiological conditions; hence, we believe that the formation of aggregation does not occur in this case. This hypothesis is confirmed by the measurement of the average hydrodynamic diameter of urease as a function of temperature (see SI-S2.1), where we observed that the size of urease remains constant up to \SI{70}{\celsius}.

Temperature-induced interfacial effects can disrupt bulk thermophoresis, as thermo-osmotic flows may arise not only around the moving object but also along substrate-solvent interfaces, influencing object drift in temperature gradients~\cite{Bregulla}. We theoretically analyze how this thermo-osmotic flow field affects the in-plane in the \textit{x}-, and the \textit{y}-directions of the urease as well as the out-of-plane in the \textit{z}-direction (see SI-S1.3). Our calculations suggest that urease can theoretically be delivered to the metamaterial at high speeds, reaching up to \SI{230}{\micro\meter\per\second} for $I_{trap}$ of \SI{1}{\milli\watt\per\square\micro\meter} (see SI-S1.3). These thermo-osmotic flows are induced without any external pressure, can be controlled by the $I_{trap}$, and are quickly switched due to the extremely fast heat conduction at these length scales. Notably, the \textit{z}-component of the thermophoretic force attracts the urease into the ASR nanocavity (Figures~\ref{Fig.2}(g) and (h)), opposing the in-plane drifts induced by thermo-osmotic flow, which assist the transportation of the urease toward the metamaterial. This competition between thermophoretic force components results in a stable trapping position that is displaced from the center of the metamaterial.

Through the manipulation of light and mass transfer at a liquid-metamaterial interface, we have demonstrated a versatile nanofluidic system on a chip that enables nano-object transportation, trapping, and manipulation. Using low trapping laser intensities, we successfully delivered and trapped single biomolecules off-center using a metamaterial tweezers. Our noninvasive optical nanotweezers approach is expected to open new avenues in nanoscience and life science research by offering an unprecedented level of control and discrimination between nano-objects without the limitations of strong-bonding fluorescent groups, thereby making single biomolecule manipulation and characterization a reality.

\section*{Materials and Methods}

\noindent \textbf{Optical trapping Setup}\

The optical tweezers consists of a  continuous-wave (CW) tunable Ti:sapphire laser tuned to \SI{930}{\nano\meter} and focused using a high numerical aperture (NA = 1.3) oil immersion objective lens (OLYMPUS UPlanFL N 100$\times$) to a spot size of $\sim$~\SI{1}{\micro\meter}. Detection of trapping events was conducted by collecting the transmitted laser light through a 50$\times$ objective lens (Nikon CF Plan), sending it to an avalanche photodiode (APD430A/M, Thorlabs), and recording it using a data acquisition card (DAQ USB-6363, NI) at a frequency of \SI{100}{\kilo\hertz}, with LabVIEW software. Given that many enzyme processes can take place on microsecond time scales~\cite{Haim} the sampling rate of \SI{100}{\kilo\hertz}, which provides a time resolution of \SI{10}{\micro\second}, is sufficient for capturing these rapid events.

\noindent \textbf{Fabrication process}\

The metamaterial consists of an array of 15 ($\it{x}$-direction) $\times$ 16 ($\it{y}$-direction) units with a period of \SI{400}{\nano\meter} in both directions and was fabricated using focused ion beam (FIB-FEI Helios G3UC) milling on a \SI{50}{\nano\meter} thin gold film (PHASIS, Geneva, BioNano) at \SI{30}{\kilo\volt} and a \SI{2}{\pico\ampere} beam current. The volume per dose was \SI{0.27}{\micro\meter\cubed\per\nano\coulomb}. The nanostructure was etched at \SI{80}{\nano\meter} depth to ensure that the structures were cut all the way through the gold film uniformly. After  fabrication, we used oxygen plasma treatment on the device for \SI{3}{\minute} - \SI{5}{\minute} to remove any gallium residue deposited on it during the milling process. The transmission and reflection spectra were obtained with a microspectrophotometer (CRAIC, 20/30 PV) with the metamaterial placed in deionized water. Then, the metamaterial was attached to a microscope cover glass with adhesive microscope spacers of \SI{10}{\micro\meter} thickness, forming a microwell. Since the focus spot has a diameter of about $\sim$\SI{1}{\micro\meter}, 3 $\times$ 3 metamolecules of the metamaterial structures were illuminated.

\noindent\textbf{Sample solutions}\

We used a solution containing heavy water, polyethylene glycol (PEG$_{6000}$, CAS No: 25322-68-3), and urease (U4002; Sigma-Aldrich). Several concentrations of urease suspended in the \SI{4.0}{\wpv} PEG$_{6000}$/D$_{2}$O water solution were used and a microwell containing urease of various concentrations was mounted and fixed on top of a piezoelectric translation stage. The role of PEG$_{6000}$ in the urease solution is to improve enzyme diffusion toward the plasmonic hotspot, to minimize enzyme overheating and denaturation under illumination with a high laser power, as well as to compensate the protein-protein interactions~\cite{WangXue}. The urease protein could then be trapped by a balance between plasmon-enhanced optical and thermophoretic forces into the ASR nano-aperture of the metamaterial. In addition, a solution of polystyrene (PS) particles with a mean diameter of \SI{20}{\nano\meter} (ThermoFisher Scientific, F8786) in \SI{4.0}{\wpv} PEG$_{6000}$/D$_{2}$O water with a concentration of \SI{0.0625}{\wpv} was used as a control. Note that the solution of each target particle was sonicated for \SI{3}{\minute}, and then \SI{8}{\micro\liter} of the final solution was placed inside the microwell on the gold film where the metamaterial was fabricated.

\noindent\textbf{Simulations}\

We used commercial finite element modeling COMSOL Multiphysics 6.2 software. A 3D model was established to solve the electromagnetic (EM) and heat transfer (HT) and (LT) problems. The modeling process was governed by a set of partial differential equations describing the full-wave EM and HT physics, and the coupling phenomena between them. The 3D EM domain was set at \SI{3.6}{\micro\metre} × \SI{3.6}{\micro\metre} × \SI{3.6}{\micro\metre} cube, consisting of an array of 7 $\times$ 7 unit cells illuminated from the substrate side by a laser with a Gaussian intensity distribution. The 7 $\times$ 7 array was simulated instead of the actual structure to reduce computational cost while still capturing the key optical behavior as well as ensuring edge effects are negligible.
The focal spot radius was set at w$_{o}$ = \SI{0.6}{\micro\metre}. Scattering boundary conditions were used on all outer EM domain boundaries. Note that the theoretical resonance is calculated at  $\lambda_{t}$ = \SI{937}{\nano\meter}, while the experimental value is measured to be $\lambda_{exp}$ = \SI{928}{\nano\meter}~\cite{Kotsifaki1,Bouloumis}. This discrepancy in both the position and magnitude of the absorption resonant peak is primarily attributed to fabrication imperfections during the focused ion beam milling process. Specifically, the edges of the ASR nano-aperture features tend to be rounded in practice, whereas simulations assume perfectly sharp edges. This deviation leads to inaccuracies in the simulated temperature response.


\bibliography{report}   

\begin{thebibliography}{10}

\bibitem{Farka}
Z.~Farka, M.~J. Mickert, M.~Pastucha, {\em et~al.}, ``Advances in optical single-molecule detection: En-route to supersensitive bioaffinity assays,'' {\em Angewandte Chemie International Edition} {\bf 59}(27), 10746--10773  (2020).

\bibitem{Silas}
S.~J. Leavesley and T.~C. Rich, ``Overcoming limitations of {FRET} measurements,'' {\em Cytometry Part A} {\bf 89}(4), 325--327  (2016).

\bibitem{Ashkin}
A.~Ashkin, J.~M. Dziedzic, J.~E. Bjorkholm, {\em et~al.}, ``Observation of a single-beam gradient force optical trap for dielectric particles,'' {\em Optics Letters} {\bf 11}(5), 288--290  (1986).

\bibitem{Bustamante}
C.~J. Bustamante, Y.~R. Chemla, S.~Liu, {\em et~al.}, ``Optical tweezers in single-molecule biophysics,'' {\em Nature Reviews Methods Primers} {\bf 1:25}, 1--25  (2021).

\bibitem{Volpe_2023}
G.~Volpe, O.~M. Maragò, H.~Rubinsztein-Dunlop, {\em et~al.}, ``Roadmap for optical tweezers,'' {\em Journal of Physics: Photonics} {\bf 5}(2), 022501  (2023).

\bibitem{Novotny}
L.~Novotny, R.~X. Bian, and X.~S. Xie, ``Theory of nanometric optical tweezers,'' {\em Physics Review Letters} {\bf 79}, 645--648  (1997).

\bibitem{Pang}
Y.~Pang and R.~Gordon, ``Optical trapping of a single protein,'' {\em Nano Letters} {\bf 12}(1), 402--406  (2012).

\bibitem{Kotsifaki1}
D.~G. Kotsifaki, V.~G. Truong, and S.~{Nic Chormaic}, ``Fano-resonant, asymmetric, metamaterial-assisted tweezers for single nanoparticle trapping,'' {\em Nano Letters} {\bf 20}(5), 3388--3395  (2020).

\bibitem{Bouloumis_2021}
T.~D. Bouloumis, D.~G. Kotsifaki, X.~Han, {\em et~al.}, ``Fast and efficient nanoparticle trapping using plasmonic connected nanoring apertures,'' {\em Nanotechnology} {\bf 32}(2), 025507  (2020).

\bibitem{Boris}
B.~Luk'yanchuk, N.~I. Zheludev, S.~A. Maier, {\em et~al.}, ``The {F}ano resonance in plasmonic nanostructures and metamaterials,'' {\em Nature Materials} {\bf 9}(9), 707--715  (2010).

\bibitem{Ahmadivand}
A.~Ahmadivand, B.~Gerislioglu, A.~Tomitaka, {\em et~al.}, ``Extreme sensitive metasensor for targeted biomarkers identification using colloidal nanoparticles-integrated plasmonic unit cells,'' {\em Biomedical Optics Express} {\bf 9}(2), 373--386  (2018).

\bibitem{KotsifakiBOE}
D.~G. Kotsifaki, R.~R. Singh, S.~{Nic Chormaic}, {\em et~al.}, ``Asymmetric split-ring plasmonic nanostructures for the optical sensing of {E}scherichia coli,'' {\em Biomedical Optics Express} {\bf 14}(9), 4875--4887  (2023).

\bibitem{Wu}
W.~Chihhui, A.~B. Khanikaev, R.~Adato, {\em et~al.}, ``Fano-resonant asymmetric metamaterials for ultrasensitive spectroscopy and identification of molecular monolayers,'' {\em Nature Materials} {\bf 11}(1), 69--75  (2012).

\bibitem{Papasimakis}
N.~Papasimakis and N.~I. Zheludev, ``Metamaterial-induced transparency: sharp {F}ano resonances and slow light,'' {\em Optics and Photonics News} {\bf 20}(10), 22--27  (2009).

\bibitem{DomnaAPL}
D.~G. Kotsifaki, V.~G. Truong, and S.~{Nic Chormaic}, ``Dynamic multiple nanoparticle trapping using metamaterial plasmonic tweezers,'' {\em Applied Physics Letters} {\bf 118}(2), 021107  (2021).

\bibitem{Bouloumis}
T.~D. Bouloumis, D.~G. Kotsifaki, and S.~{Nic Chormaic}, ``Enabling self-induced back-action trapping of gold nanoparticles in metamaterial plasmonic tweezers,'' {\em Nano Letters} {\bf 23}(11), 4723--4731  (2023).

\bibitem{Baffou_Nat}
G.~Baffou, F.~Cichos, and R.~Quidant, ``Applications and challenges of thermoplasmonics,'' {\em Nature Materials} {\bf 19}, 946--958  (2020).

\bibitem{Zijlstra_NatNanotech}
P.~Zijlstra, P.~M.~R. Paulo, and M.~Orrit, ``Optical detection of single non-absorbing molecules using the surface plasmon resonance of a gold nanorod,'' {\em Nature Nanotechnology} {\bf 7}, 379--382  (2012).

\bibitem{Wurger_2010}
A.~Würger, ``Thermal non-equilibrium transport in colloids,'' {\em Reports on Progress in Physics} {\bf 73}(12), 126601  (2010).

\bibitem{Bregulla}
A.~P. Bregulla, A.~W\"urger, K.~G\"unther, {\em et~al.}, ``Thermo-osmotic flow in thin films,'' {\em Physics Review Letters} {\bf 116}, 188303  (2016).

\bibitem{Weinert}
F.~M. Weinert and D.~Braun, ``Optically driven fluid flow along arbitrary microscale patterns using thermoviscous expansion,'' {\em Journal of Applied Physics} {\bf 104}(10), 104701  (2008).

\bibitem{Sano}
H.-R. Jiang, H.~Wada, N.~Yoshinaga, {\em et~al.}, ``Manipulation of colloids by a nonequilibrium depletion force in a temperature gradient,'' {\em Physics Review Letters} {\bf 102}, 208301  (2009).

\bibitem{Jiang_2024}
Z.~Jiang, Y.~Sun, Y.~Gao, {\em et~al.}, ``Fast lipid vesicles and dielectric particles migration using thermal-gradient-induced forces,'' {\em Journal of Optics} {\bf 26}(9), 095301  (2024).

\bibitem{Franz_2019}
M.~Fränzl, T.~Thalheim, J.~Adler, {\em et~al.}, ``Thermophoretic trap for single amyloid fibril and protein aggregation studies,'' {\em Nature Methods} {\bf 16}, 611--614  (2019).

\bibitem{Follmer}
C.~Follmer, F.~V. Pereira, N.~P. da~Silveira, {\em et~al.}, ``Jack bean urease (ec 3.5.1.5) aggregation monitored by dynamic and static light scattering,'' {\em Biophysical Chemistry} {\bf 111}, 79--87  (2004).

\bibitem{Damien}
D.~A.~F. Lynch, N.~P. Mapstone, F.~Lewis, {\em et~al.}, ``Serum and gastric luminal epidermal growth factor in {H}elicobacter pylori—associated gastritis and peptic ulcer disease,'' {\em Helicobacter} {\bf 1}(4), 219--226  (1996).

\bibitem{Pundir}
C.~Pundir, S.~Jakhar, and V.~Narwal, ``Determination of urea with special emphasis on biosensors: A review,'' {\em Biosensors and Bioelectronics} {\bf 123}, 36--50  (2019).

\bibitem{Gupta}
V.~Gupta, S.~Bhavanasi, M.~Quadir, {\em et~al.}, ``Protein {PEG}ylation for cancer therapy: bench to bedside,'' {\em Journal of Cell Communication and Signaling} {\bf 13}, 319--330  (2019).

\bibitem{BALASUBRAMANIAN2010274}
A.~Balasubramanian and K.~Ponnuraj, ``Crystal structure of the first plant urease from {J}ack bean: 83 years of journey from its first crystal to molecular structure,'' {\em Journal of Molecular Biology} {\bf 400}(3), 274--283  (2010).

\bibitem{Quinn2025}
D.~J. Quinn, D.~Paul, and F.~Cichos, ``{Thermofluidic Nonequilibrium Assembly of Reconfigurable Functional Structures},'' {\em ACS Nano} {\bf 19}(23), 21820--21829  (2025).

\bibitem{Iacopini_2003}
S.~Iacopini and R.~Piazza, ``Thermophoresis in protein solutions,'' {\em Europhysics Letters} {\bf 63}(2), 247  (2003).

\bibitem{Putnam}
S.~A. Putnam, D.~G. Cahill, and G.~C.~L. Wong, ``Temperature dependence of thermodiffusion in aqueous suspensions of charged nanoparticles,'' {\em Langmuir} {\bf 23}(18), 9221--9228  (2007).

\bibitem{Pu}
D.~Pu, A.~Panahi, G.~Natale, {\em et~al.}, ``A mode-coupling model of colloid thermophoresis in aqueous systems: Temperature and size dependencies of the {S}oret coefficient,'' {\em Nano Letters} {\bf 24}(9), 2798--2804  (2024).

\bibitem{Young}
G.~Young, N.~Hundt, D.~Cole, {\em et~al.}, ``Quantitative mass imaging of single biological macromolecules,'' {\em Science} {\bf 360}(6387), 423--427  (2018).

\bibitem{Madan}
L.~K. Madan, C.~L. Welsh, A.~P. Kornev, {\em et~al.}, ``{The “violin model”: Looking at community networks for dynamic allostery},'' {\em The Journal of Chemical Physics} {\bf 158}(8), 081001  (2023).

\bibitem{Feder}
M.~J. Feder, A.~Akyel, V.~J. Morasko, {\em et~al.}, ``Temperature-dependent inactivation and catalysis rates of plant-based ureases for engineered biomineralization,'' {\em Engineering Reports} {\bf 3}(2), e12299  (2021).

\bibitem{OMAR}
S.~Omar and M.~Beauregard, ``Dissociation and unfolding of jack bean urease studied by fluorescence emission spectroscopy,'' {\em Journal of Biotechnology} {\bf 39}(3), 221--228  (1995).

\bibitem{OMAR1994}
S.~Omar and M.~Beauregard, ``Detection of $\alpha$-urease dissociation by fluorescence emission spectroscopy,'' {\em Biochemical and Biophysical Research Communications} {\bf 201}(3), 1096--1099  (1994).

\bibitem{Grancic}
P.~Grancic, V.~Illeova, M.~Polakovic, {\em et~al.}, ``Thermally induced inactivation and aggregation of urease: Experiments and population balance modelling,'' {\em Chemical Engineering Science} {\bf 70}, 14--21  (2012).

\bibitem{Vardar}
V.~Gokay, A.~Azade, and Y.~Melda~Altikatoglou, ``Development of urea biosensor using non-covalent complexes of urease with aldehyde derivative of peg and analysis on serum samples,'' {\em Preparative Biochemistry \& Biotechnology} {\bf 49}(9), 868--875  (2019).

\bibitem{Haim}
H.~Y. Aviram, M.~Pirchi, H.~Mazal, {\em et~al.}, ``Direct observation of ultrafast large-scale dynamics of an enzyme under turnover conditions,'' {\em Proceedings of the National Academy of Sciences} {\bf 115}(13), 3243--3248  (2018).

\bibitem{WangXue}
X.~Wang, J.~Bowman, S.~Tu, {\em et~al.}, ``Polyethylene glycol crowder’s effect on enzyme aggregation, thermal stability, and residual catalytic activity,'' {\em Langmuir} {\bf 37}(28), 8474--8485  (2021).

\end{thebibliography}
\bibliographystyle{spiejour}   


\noindent{\textbf{Acknowledgments}}\\
The authors thank M. Ozer for technical assistance, T. Bouloumis for  device fabrication, N. Ishizu from the Engineering Section at Okinawa Institute of Science and Technology Graduate University (OIST), and P. Puchenkov from the Scientific Computing and Data Analysis Section at OIST. SNC acknowledges early discussions with F.A. Samatey and M.A. Price on protein structure and detection methods.

\noindent\textbf{Funding} \\
This work was supported by funding from OIST Graduate University. DGK acknowledges DKU Startup Fund, DKU Summer Research Program Grant 2024, and support from the Sumitomo Foundation Grant for Basic Science Research Project. VGT acknowledges support from JSPS KAKENHI, Grant-in-Aid for Scientific Research (C), Project Number JP23K04618.

\noindent \textbf{Author contributions}\\
SNC, DGK and VGT conceived the project. SNC initiated and supervised the project. VGT designed and carried out all the numerical simulations. FC proposed conceptions and interpretations for the numerical simulations. DGK designed and performed all the experiments and analyzed the experimental data. MD conducted the urease activity measurements. DGK and VGT wrote the original draft and contributed equally to this work. All authors discussed the results, and reviewed and edited the manuscript.

\noindent \textbf{Competing interests}\\
The authors declare that they have no competing interests.

\noindent\textbf{Data and materials availability}\\
All data needed to evaluate the conclusions in the paper are present in the paper and/or the Supplementary Materials. Additional data related to this paper may be requested from the authors.

\noindent\textbf{Additional information}\\
Supplementary information available at...

\end{spacing}
\end{document}